\documentclass[showpacs,aps,prd]{revtex4}
\input epsf

\begin{document}

\title{Deciphering triply heavy baryons in terms of QCD sum rules}
\author{Jian-Rong Zhang and Ming-Qiu Huang}
\affiliation{Department of Physics, National University of Defense
Technology, Hunan 410073, China}

\begin{abstract}
The mass spectra of ground-state triply heavy baryons are
systematically unscrambled and computed in QCD sum rules. With a
tentative $(QQ)-(Q')$ configuration for $QQQ'$, the interpolating
currents representing the triply heavy baryons are proposed.
Technically, contributions of the operators up to dimension six are
included in operator product expansion (OPE). The numerical results
are presented in comparison with other theoretical predictions.
\end{abstract}
\maketitle
\section{Introduction}\label{sec1}
The triply heavy baryon, wherever light quarks are absent, is well
and truly not a new topic but with a history. As one refers to the
studies on their properties, it can be traced back to two more
decades ago \cite{Hasenfratz,Bjorken}. However, contrasted with the
singly and doubly heavy baryons (such as Refs. \cite{Bagan,heavy
baryons}), only infrequent attention has been paid to the triply
heavy baryons. Whereas, the case for triply heavy baryon may be
improved and several approaches have already appeared in recent
years, such as effective field theory \cite{EFT}, lattice QCD
\cite{lat}, variational method \cite{YuJia}, bag model
\cite{renewed1}, quark models \cite{quark model,Martynenko,renewed2}
etc., for which is gradually becoming a exciting and remarkable
theme nowadays. First, the field of heavy hadron spectroscopy is
experiencing a rapid advancement, which is mainly propelled by the
continuous discovery of hadronic resonances (for reviews, e.g., see
\cite{overview1,overview2}). While experimentally reconstructing a
candidate for $\Omega_{ccc}$ is very difficult, it is not
unthinkable according to Ref. \cite{Bjorken}. Especially for the
startup of Large Hadron Collider, it seems rather promising to
establish triply heavy baryons in future
\cite{production1,production2,production3}. Second, investigation of
the triply heavy baryon is of great interest in understanding the
dynamics of QCD at the hadronic scale. Although the statement that
QCD is the correct theory underlying strong interaction has been
commonly accepted and QCD is simple and elegant in its formulation,
many questions concerning dynamics of the quarks and gluons at large
distances remain unanswered or, at most, understood only at a
qualitative level. The quantitative description of the hadronic
properties runs into however arduous difficulties. For example, it
is a great challenge to extract information on the spectrum from the
rather simple Lagrangian of QCD. That's because low energy QCD
involves a regime where it is futile to attempt perturbative
calculations and, inevitably, one has to treat a genuinely strong
field in nonperturbative methods. Briefly recapitulating the second
point, triply heavy baryons, free of light quark contamination, are
ideal prototypes to refine one's present understanding of heavy
quark dynamics and may serve as a clean probe to the interplay
between perturbative and nonperturbative QCD. Also stimulated by the
above two aspects, it is interesting and significative to study
their properties like masses through nonperturbative approaches, and
the practitioner may resort to a vigorous and reliable working tool
in hadron physics, the QCD sum rule \cite{svzsum}, which is a
nonperturbative analytic formalism firmly entrenched in QCD. On the
sum rule analysis, the triply heavy baryon systems are analogous to
the cases of charmonium and bottomonium, where light quarks are all
absent. In fact, there have already been some works on calculating
the charmonium and bottomonium masses in QCD sum rules, such as
\cite{charmonium and bottomonium}. The $c$ and $b$ quark masses can
also been determined from considering the two-point correlation
function of the $\bar{Q}\gamma_{\mu}Q$ current ($Q=c~\mbox{or}~b$),
for instance in \cite{svzsum,c and b}, and some impressive
progresses were made in updating the values of $m_{c}$ and $m_{b}$
later, including the $O(\alpha_{s}^{2})$ perturbative corrections
\cite{alpha2}. In addition, the semileptonic decays of $B_{c}$ have
been investigated by three-point sum rules \cite{Bc}. Thereby, it is
feasible for QCD sum rules to study triply heavy baryons and we
would like to carry out the sum rule calculations of their spectra
in this work.

The paper is organized as follows. In Sec. \ref{sec2}, QCD sum rules
for the triply heavy baryons are introduced, and both the
phenomenological representation and QCD side are derived, followed
by the numerical analysis to extract the spectra and a comparison
with other theoretical calculations in Sec. \ref{sec3}. Section
\ref{sec4} contains a brief summary and outlook.

\section{Triply heavy baryon QCD sum rules}\label{sec2}
A generic QCD sum rule calculation consists of three main
ingredients: an approximate description of the correlator in terms
of intermediate states through the dispersion relation, a
description of the same correlation function in terms of QCD degrees
of freedom via an OPE, and a procedure for matching these two
descriptions and extracting the parameters that characterize the
hadronic state of interest. Concretely, coming down to the mass sum
rules for triply heavy baryon $QQQ'$ (here $Q$ and $Q'$ can be the
same or differently heavy quarks, $c$ or $b$), the starting point is
the two-point correlation function
\begin{eqnarray}\label{correlator}
\Pi(q^{2})=i\int
d^{4}x\mbox{e}^{iq.x}\langle0|T[j(x)\overline{j}(0)]|0\rangle.
\end{eqnarray}
Lorentz covariance implies that the correlation function
(\ref{correlator}) has the form
\begin{eqnarray}
\Pi(q^{2})=\rlap/q\Pi_{1}(q^{2})+\Pi_{2}(q^{2}).
\end{eqnarray}
For each invariant function $\Pi_{1}$ and $\Pi_{2}$, a sum rule can
be obtained.

In the phenomenological side, the correlator can be expressed as a
dispersion integral over a physical spectral function
\begin{eqnarray}
\Pi(q^{2})=\lambda^{2}_H\frac{\rlap/q+M_{H}}{M_{H}^{2}-q^{2}}+\frac{1}{\pi}\int_{s_{0}}
^{\infty}ds\frac{\mbox{Im}\Pi^{\mbox{phen}}(s)}{s-q^{2}}+\mbox{subtractions},
\end{eqnarray}
where $M_{H}$ denotes the mass of the triply heavy baryon. In
obtaining the above expression, the Dirac and Rarita-Schwinger
spinor sum relations,
\begin{eqnarray}
\sum_{s}N(q,s)\bar{N}(q,s)=\rlap/q+M_{H},
\end{eqnarray}
for spin-$\frac{1}{2}$ baryon, and
\begin{eqnarray}
\sum_{s}N_{\mu}(q,s)\bar{N}_{\nu}(q,s)=(\rlap/q+M_{H})(g_{\mu\nu}-\frac{1}{3}\gamma_{\mu}\gamma_{\nu}+\frac{q_{\mu}\gamma_{\nu}-q_{\nu}\gamma_{\mu}}{3M_{H}}-\frac{2q_{\mu}q_{\nu}}{3M_{H}^{2}}),
\end{eqnarray}
for spin-$\frac{3}{2}$ baryon, have been used.

In the OPE side, the correlation function can be written in terms of
a dispersion relation as
\begin{eqnarray}
\Pi_{i}(q^{2})=\int_{(2m_{Q}+m_{Q'})^{2}}^{\infty}ds\frac{\rho_{i}(s)}{s-q^{2}},~~i=1,2,
\end{eqnarray}
where the spectral density is given by the imaginary part of the
correlation function
\begin{eqnarray}
\rho_{i}(s)=\frac{1}{\pi}\mbox{Im}\Pi_{i}^{\mbox{OPE}}(s).
\end{eqnarray}
In detail, the spectral densities are calculated and embodied in
Sec. \ref{densities}.

After equating the two sides, assuming quark-hadron duality, and
making a Borel transform, the sum rules can be written as
\begin{eqnarray}
\lambda_{H}^{2}e^{-M_{H}^{2}/M^{2}}&=&\int_{(2m_{Q}+m_{Q'})^{2}}^{s_{0}}ds\rho_{1}(s)e^{-s/M^{2}},
\end{eqnarray}
\begin{eqnarray}
\lambda_{H}^{2}M_{H}e^{-M_{H}^{2}/M^{2}}&=&\int_{(2m_{Q}+m_{Q'})^{2}}^{s_{0}}ds\rho_{2}(s)e^{-s/M^{2}}.
\end{eqnarray}
To eliminate the baryon coupling constant $\lambda_H$, one reckons
the ratio of derivative of the sum rule and itself and yields
\begin{eqnarray}\label{sum rule q}
M_{H}^{2}&=&\int_{(2m_{Q}+m_{Q'})^{2}}^{s_{0}}ds\rho_{1}(s)s
e^{-s/M^{2}}/
\int_{(2m_{Q}+m_{Q'})^{2}}^{s_{0}}ds\rho_{1}(s)e^{-s/M^{2}},
\end{eqnarray}
\begin{eqnarray}\label{sum rule m}
M_{H}^{2}&=&\int_{(2m_{Q}+m_{Q'})^{2}}^{s_{0}}ds\rho_{2}(s)s
e^{-s/M^{2}}/
\int_{(2m_{Q}+m_{Q'})^{2}}^{s_{0}}ds\rho_{2}(s)e^{-s/M^{2}}.
\end{eqnarray}
\subsection{The interpolating currents}
In a tentative picture for $Q Q Q'$ system, the $Q'$ orbits the
bound $Q Q$ pair. The $(QQ)-(Q')$ structure may be described similar
to $\bar{Q}Q'$ mesons, where the $Q Q$ pair plays the same role of
the antiquark $\bar{Q}$ in $\bar{Q}Q'$. The study of such
configuration can help one to adopt the appropriate interpolating
currents. For the ground states, the currents are correlated with
the spin-parity quantum numbers $0^{+}$ and $1^{+}$ for the heavy $Q
Q$ diquark system, along with the quark $Q'$ forming the state with
$J^{P}=\frac{1}{2}^{+}$ and the pair of degenerate states. For the
latter case, the $Q Q$ diquark has spin $1$, and the spin of the
third quark is either parallel, $J^{P}=\frac{3}{2}^{+}$, or
antiparallel, $J^{P}=\frac{1}{2}^{+}$, to the diquark.  The choice
of $\Gamma_{k}$ and $\Gamma_{k}^{'}$ matrices in baryonic currents
may be determined according to the rules in \cite{evs}. For the
baryon with $J^{P}=\frac{3}{2}^{+}$, the current may be gained using
$SU(3)$ symmetry relations \cite{Ioffe}. Consequently, following
forms of currents are adopted
\begin{eqnarray}
j_{\Omega_{QQQ}}&=&\varepsilon_{abc}(Q_{a}^{T}C\Gamma_{k}Q_{b})\Gamma_{k}^{'}Q_{c},\nonumber\\
j_{\Omega_{QQQ'}}&=&\varepsilon_{abc}(Q_{a}^{T}C\Gamma_{k}Q_{b})\Gamma_{k}^{'}Q'_{c},\\
j_{\Omega_{QQQ'}^{*}}&=&\varepsilon_{abc}\frac{1}{\sqrt{3}}[2(Q_{a}^{T}C\Gamma_{k}Q'_{b})\Gamma_{k}^{'}Q_{c}+(Q_{a}^{T}C\Gamma_{k}Q_{b})\Gamma_{k}^{'}Q'_{c}],\nonumber\\
j_{\Omega_{QQQ'}'}&=&\varepsilon_{abc}(Q_{a}^{T}C\Gamma_{k}Q_{b})\Gamma_{k}^{'}Q'_{c},\nonumber
\end{eqnarray}
Here the index $T$ means matrix transposition, $C$ is the charge
conjugation matrix, $a$, $b$, and $c$ are color indices, with $Q$
and $Q'$ denote heavy quarks. The categories of ground-state triply
heavy baryons  and the choice of $\Gamma_{k}$ and $\Gamma_{k}^{'}$
matrices are listed in TABLE \ref{table:1}.
\begin{table}[htb!]\caption{The choice of $\Gamma_{k}$ and $\Gamma_{k}^{'}$ matrices in baryonic currents. The index $d$ in $S_{d}$, $L_{d}$, and $J_{d}^{P_{d}}$ means diquark. $\{QQ\}$ denotes the diquark in the axial
vector state and $[QQ]$ denotes diquark in the scalar state.}
 \centerline{\begin{tabular}{ p{1.5cm} p{2.5cm} p{2.0cm} p{1cm} p{1cm} p{2.0cm} p{2.0cm} p{2.0cm}} \hline\hline
Baryon              & quark content        &$J^{P}$               &  $S_{d}$     &  $L_{d}$     &  $J_{d}^{P_{d}}$         &   $\Gamma_{k}$     &     $\Gamma_{k}^{'}$           \\
$\Omega_{QQQ}$      &$\{QQ\}Q$             &$\frac{3}{2}^{+} $    &      1       &      0       &        $1^{+}$           &   $\gamma_{\mu}$   &     $1$                        \\
\hline
$\Omega_{QQQ'}$     &$\{QQ\}Q'$            &$\frac{1}{2}^{+} $    &      1       &      0       &        $1^{+}$           &   $\gamma_{\mu}$   &     $\gamma_{\mu}\gamma_{5}$   \\
\hline
$\Omega_{QQQ'}^{*}$ &$\{QQ\}Q'$            &$\frac{3}{2}^{+} $    &      1       &      0       &        $1^{+}$           &   $\gamma_{\mu}$   &     $1$                        \\
\hline
$\Omega_{QQQ'}'$    &$[QQ]Q'$              &$\frac{1}{2}^{+} $    &      0       &      0       &        $0^{+}$           &   $\gamma_{5}$     &     $1$                        \\
\hline\hline
\end{tabular}}
\label{table:1}
\end{table}
\subsection{The spectra densities}\label{densities}
Implementing the calculation of the OPE side, we work at leading
order in $\alpha_{s}$ and consider condensates up to dimension six.
To keep the heavy-quark mass finite, one uses the momentum-space
expression for the heavy-quark propagator. The final result is
dimensionally regularized at $D=4$. It should be distinguished for
spectral densities of two sort triply heavy baryons, namely,
containing the same heavy quark or differently. First, with
\begin{eqnarray}
\rho_{1}(s)&=&\frac{3}{2^{3}\pi^{4}}\int_{\alpha_{min}}^{\alpha_{max}}d\alpha\{\int_{0}^{\beta_{1}}d\beta+\int_{\beta_{2}}^{\frac{1}{\alpha}}d\beta\}\frac{\alpha^{2}\beta^{2}(1-\alpha\beta)^{2}}{(\alpha+\beta)^{4}}[\alpha\beta(1-\alpha\beta)
s-(\alpha^{2}+\beta^{2}+\alpha\beta+1)m_{Q}^{2}]s\nonumber\\&&{}
+\frac{3^{2}}{2^{4}\pi^{4}}\int_{\alpha_{min}}^{\alpha_{max}}d\alpha\{\int_{0}^{\beta_{1}}d\beta+\int_{\beta_{2}}^{\frac{1}{\alpha}}d\beta\}\frac{\alpha\beta(1-\alpha\beta)}{(\alpha+\beta)^{4}}[\alpha\beta(1-\alpha\beta)
s-(\alpha^{2}+\beta^{2}+\alpha\beta+1)m_{Q}^{2}]^{2}\nonumber\\&&{}
+\frac{3}{2^{2}\pi^{4}}m_{Q}^{2}\int_{\alpha_{min}}^{\alpha_{max}}d\alpha\{\int_{0}^{\beta_{1}}d\beta+\int_{\beta_{2}}^{\frac{1}{\alpha}}d\beta\}\frac{\alpha\beta}{(\alpha+\beta)^{2}}[\alpha\beta(1-\alpha\beta)
s-(\alpha^{2}+\beta^{2}+\alpha\beta+1)m_{Q}^{2}]\nonumber\\&&{}
+\frac{3\langle
g^{2}G^{2}\rangle}{2^{6}\pi^{4}}\int_{\alpha_{min}}^{\alpha_{max}}d\alpha\{\int_{0}^{\beta_{1}}d\beta+\int_{\beta_{2}}^{\frac{1}{\alpha}}d\beta\}\frac{\alpha^{2}\beta^{2}(1-\alpha\beta)}{(\alpha+\beta)^{2}},\nonumber\\
\rho_{2}(s)&=&\frac{3}{2^{3}\pi^{4}}m_{Q}\int_{\alpha_{min}}^{\alpha_{max}}d\alpha\{\int_{0}^{\beta_{1}}d\beta+\int_{\beta_{2}}^{\frac{1}{\alpha}}d\beta\}\frac{\alpha\beta(1-\alpha\beta)^{2}}{(\alpha+\beta)^{4}}[\alpha\beta(1-\alpha\beta)
s-(\alpha^{2}+\beta^{2}+\alpha\beta+1)m_{Q}^{2}]s\nonumber\\&&{}
+\frac{3}{2^{3}\pi^{4}}m_{Q}\int_{\alpha_{min}}^{\alpha_{max}}d\alpha\{\int_{0}^{\beta_{1}}d\beta+\int_{\beta_{2}}^{\frac{1}{\alpha}}d\beta\}\frac{1-\alpha\beta}{(\alpha+\beta)^{4}}[\alpha\beta(1-\alpha\beta)
s-(\alpha^{2}+\beta^{2}+\alpha\beta+1)m_{Q}^{2}]^{2}\nonumber\\&&{}
+\frac{3}{2^{2}\pi^{4}}m_{Q}^{3}\int_{\alpha_{min}}^{\alpha_{max}}d\alpha\{\int_{0}^{\beta_{1}}d\beta+\int_{\beta_{2}}^{\frac{1}{\alpha}}d\beta\}\frac{1}{(\alpha+\beta)^{2}}[\alpha\beta(1-\alpha\beta)
s-(\alpha^{2}+\beta^{2}+\alpha\beta+1)m_{Q}^{2}]\nonumber\\&&{}
+\frac{\langle
g^{2}G^{2}\rangle}{2^{5}\pi^{4}}m_{Q}\int_{\alpha_{min}}^{\alpha_{max}}d\alpha\{\int_{0}^{\beta_{1}}d\beta+\int_{\beta_{2}}^{\frac{1}{\alpha}}d\beta\}\frac{(1-\alpha\beta)[(1-\alpha\beta)^{2}-\alpha\beta(\alpha+\beta)^{2}]}{(\alpha+\beta)^{4}},\nonumber
\end{eqnarray}
for $\Omega_{QQQ}$. The integration limits are given by
\begin{eqnarray}
\alpha_{min}=\sqrt{\frac{(s^{2}-6m_{Q}^{2}s-3m_{Q}^{4})-(s-m_{Q}^{2})\sqrt{(s-m_{Q}^{2})(s-9m_{Q}^{2})}}{8m_{Q}^{2}s}},\nonumber\\
\alpha_{max}=\sqrt{\frac{(s^{2}-6m_{Q}^{2}s-3m_{Q}^{4})+(s-m_{Q}^{2})\sqrt{(s-m_{Q}^{2})(s-9m_{Q}^{2})}}{8m_{Q}^{2}s}},\nonumber\\
\beta_{1}=\frac{\alpha(s-m_{Q}^{2})-\sqrt{\alpha^{2}s^{2}-6m_{Q}^{2}\alpha^{2}s-4m_{Q}^{4}-3m_{Q}^{4}\alpha^{2}-4m_{Q}^{2}\alpha^{4}s}}{2(\alpha^{2}
s+m_{Q}^{2})}, \mbox{and} \nonumber\\
\beta_{2}=\frac{\alpha(s-m_{Q}^{2})+\sqrt{\alpha^{2}s^{2}-6m_{Q}^{2}\alpha^{2}s-4m_{Q}^{4}-3m_{Q}^{4}\alpha^{2}-4m_{Q}^{2}\alpha^{4}s}}{2(\alpha^{2}
s+m_{Q}^{2})}.\nonumber
\end{eqnarray}
Next, with
\begin{eqnarray}
\rho_{1}(s)&=&-\frac{3}{2^{2}\pi^{4}}\int_{\alpha_{min}}^{\alpha_{max}}d\alpha\{\int_{0}^{\beta_{1}}d\beta+\int_{\beta_{2}}^{\frac{1}{\alpha}}d\beta\}\frac{\alpha^{2}\beta^{2}(1-\alpha\beta)^{2}}{(\alpha+\beta)^{4}}[\alpha\beta(1-\alpha\beta)
s-((\alpha+\beta)^{2}m_{Q}^{2}+(1-\alpha\beta)m_{Q'}^{2})]s\nonumber\\&&{}
-\frac{3^{2}}{2^{3}\pi^{4}}\int_{\alpha_{min}}^{\alpha_{max}}d\alpha\{\int_{0}^{\beta_{1}}d\beta+\int_{\beta_{2}}^{\frac{1}{\alpha}}d\beta\}\frac{\alpha\beta(1-\alpha\beta)}{(\alpha+\beta)^{4}}[\alpha\beta(1-\alpha\beta)
s-((\alpha+\beta)^{2}m_{Q}^{2}+(1-\alpha\beta)m_{Q'}^{2})]^{2}\nonumber\\&&{}
-\frac{3}{2\pi^{4}}m_{Q}^{2}\int_{\alpha_{min}}^{\alpha_{max}}d\alpha\{\int_{0}^{\beta_{1}}d\beta+\int_{\beta_{2}}^{\frac{1}{\alpha}}d\beta\}\frac{\alpha\beta}{(\alpha+\beta)^{2}}[\alpha\beta(1-\alpha\beta)
s-((\alpha+\beta)^{2}m_{Q}^{2}+(1-\alpha\beta)m_{Q'}^{2})]\nonumber\\&&{}
-\frac{3\langle
g^{2}G^{2}\rangle}{2^{5}\pi^{4}}\int_{\alpha_{min}}^{\alpha_{max}}d\alpha\{\int_{0}^{\beta_{1}}d\beta+\int_{\beta_{2}}^{\frac{1}{\alpha}}d\beta\}\frac{\alpha^{2}\beta^{2}(1-\alpha\beta)}{(\alpha+\beta)^{2}},\nonumber\\
\rho_{2}(s)&=&-\frac{3}{2\pi^{4}}m_{Q'}\int_{\alpha_{min}}^{\alpha_{max}}d\alpha\{\int_{0}^{\beta_{1}}d\beta+\int_{\beta_{2}}^{\frac{1}{\alpha}}d\beta\}\frac{\alpha\beta(1-\alpha\beta)^{2}}{(\alpha+\beta)^{4}}[\alpha\beta(1-\alpha\beta)
s-((\alpha+\beta)^{2}m_{Q}^{2}+(1-\alpha\beta)m_{Q'}^{2})]s\nonumber\\&&{}
-\frac{3}{2\pi^{4}}m_{Q'}\int_{\alpha_{min}}^{\alpha_{max}}d\alpha\{\int_{0}^{\beta_{1}}d\beta+\int_{\beta_{2}}^{\frac{1}{\alpha}}d\beta\}\frac{1-\alpha\beta}{(\alpha+\beta)^{4}}[\alpha\beta(1-\alpha\beta)
s-((\alpha+\beta)^{2}m_{Q}^{2}+(1-\alpha\beta)m_{Q'}^{2})]^{2}\nonumber\\&&{}
-\frac{3}{\pi^{4}}m_{Q}^{2}m_{Q'}\int_{\alpha_{min}}^{\alpha_{max}}d\alpha\{\int_{0}^{\beta_{1}}d\beta+\int_{\beta_{2}}^{\frac{1}{\alpha}}d\beta\}\frac{1}{(\alpha+\beta)^{2}}[\alpha\beta(1-\alpha\beta)
s-((\alpha+\beta)^{2}m_{Q}^{2}+(1-\alpha\beta)m_{Q'}^{2})]\nonumber\\&&{}
-\frac{\langle
g^{2}G^{2}\rangle}{2^{3}\pi^{4}}m_{Q'}\int_{\alpha_{min}}^{\alpha_{max}}d\alpha\{\int_{0}^{\beta_{1}}d\beta+\int_{\beta_{2}}^{\frac{1}{\alpha}}d\beta\}\frac{(1-\alpha\beta)[(1-\alpha\beta)^{2}-\alpha\beta(\alpha+\beta)^{2}]}{(\alpha+\beta)^{4}},\nonumber
\end{eqnarray}
for $\Omega_{QQQ'}$,
\begin{eqnarray}
\rho_{1}(s)&=&\frac{1}{2^{3}\pi^{4}}\int_{\alpha_{min}}^{\alpha_{max}}d\alpha\{\int_{0}^{\beta_{1}}d\beta+\int_{\beta_{2}}^{\frac{1}{\alpha}}d\beta\}\frac{\alpha^{2}\beta^{2}(1-\alpha\beta)^{2}}{(\alpha+\beta)^{4}(\alpha^{2}+1)}[4(\alpha+\beta)^{2}\nonumber\\&&{}
+(\alpha^{2}+1)][\alpha\beta(1-\alpha\beta)s-((\alpha+\beta)^{2}m_{Q}^{2}+(1-\alpha\beta)m_{Q'}^{2})]s\nonumber\\&&{}
+\frac{3}{2^{4}\pi^{4}}\int_{\alpha_{min}}^{\alpha_{max}}d\alpha\{\int_{0}^{\beta_{1}}d\beta+\int_{\beta_{2}}^{\frac{1}{\alpha}}d\beta\}\frac{\alpha\beta(1-\alpha\beta)}{(\alpha+\beta)^{4}(\alpha^{2}+1)}[4(\alpha+\beta)^{2}\nonumber\\&&{}
+(\alpha^{2}+1)][\alpha\beta(1-\alpha\beta)s-((\alpha+\beta)^{2}m_{Q}^{2}+(1-\alpha\beta)m_{Q'}^{2})]^{2}\nonumber\\&&{}
+\frac{1}{2^{2}\pi^{4}}\int_{\alpha_{min}}^{\alpha_{max}}d\alpha\{\int_{0}^{\beta_{1}}d\beta+\int_{\beta_{2}}^{\frac{1}{\alpha}}d\beta\}\frac{\alpha}{(\alpha+\beta)^{2}(\alpha^{2}+1)}[4(1-\alpha\beta)(\alpha+\beta)m_{Q}m_{Q'}\nonumber\\&&{}
+\beta(\alpha^{2}+1)m_{Q}^{2}][\alpha\beta(1-\alpha\beta)
s-((\alpha+\beta)^{2}m_{Q}^{2}+(1-\alpha\beta)m_{Q'}^{2})]\nonumber\\&&{}
+\frac{\langle
g^{2}G^{2}\rangle}{2^{6}\pi^{4}}\int_{\alpha_{min}}^{\alpha_{max}}d\alpha\{\int_{0}^{\beta_{1}}d\beta+\int_{\beta_{2}}^{\frac{1}{\alpha}}d\beta\}\frac{\alpha^{2}\beta(1-\alpha\beta)}{(\alpha+\beta)^{2}(\alpha^{2}+1)}[4(1-\alpha\beta)(\alpha+\beta)+\beta(\alpha^{2}+1)],\nonumber
\end{eqnarray}
\begin{eqnarray}
\rho_{2}(s)&=&\frac{1}{2^{3}\pi^{4}}\int_{\alpha_{min}}^{\alpha_{max}}d\alpha\{\int_{0}^{\beta_{1}}d\beta+\int_{\beta_{2}}^{\frac{1}{\alpha}}d\beta\}\frac{\alpha\beta(1-\alpha\beta)}{(\alpha+\beta)^{4}(\alpha^{2}+1)}[4\beta(\alpha+\beta)^{3}m_{Q}\nonumber\\&&{}
+(1-\alpha\beta)(\alpha^{2}+1)m_{Q'}][\alpha\beta(1-\alpha\beta)
s-((\alpha+\beta)^{2}m_{Q}^{2}+(1-\alpha\beta)m_{Q'}^{2})]s\nonumber\\&&{}
+\frac{1}{2^{3}\pi^{4}}\int_{\alpha_{min}}^{\alpha_{max}}d\alpha\{\int_{0}^{\beta_{1}}d\beta+\int_{\beta_{2}}^{\frac{1}{\alpha}}d\beta\}\frac{1}{(\alpha+\beta)^{4}(\alpha^{2}+1)}[4\beta(\alpha+\beta)^{3}m_{Q}\nonumber\\&&{}
+(1-\alpha\beta)(\alpha^{2}+1)m_{Q'}][\alpha\beta(1-\alpha\beta)
s-((\alpha+\beta)^{2}m_{Q}^{2}+(1-\alpha\beta)m_{Q'}^{2})]^{2}\nonumber\\&&{}
+\frac{1}{2^{2}\pi^{4}}m_{Q}^{2}m_{Q'}\int_{\alpha_{min}}^{\alpha_{max}}d\alpha\{\int_{0}^{\beta_{1}}d\beta+\int_{\beta_{2}}^{\frac{1}{\alpha}}d\beta\}\frac{1}{(\alpha+\beta)^{2}(\alpha^{2}+1)}[4(\alpha+\beta)^{2}\nonumber\\&&{}
+(\alpha^{2}+1)][\alpha\beta(1-\alpha\beta)
s-((\alpha+\beta)^{2}m_{Q}^{2}+(1-\alpha\beta)m_{Q'}^{2})]\nonumber\\&&{}
+\frac{\langle
g^{2}G^{2}\rangle}{3\cdot2^{5}\pi^{4}}\int_{\alpha_{min}}^{\alpha_{max}}d\alpha\{\int_{0}^{\beta_{1}}d\beta+\int_{\beta_{2}}^{\frac{1}{\alpha}}d\beta\}\frac{1}{(\alpha+\beta)^{4}(\alpha^{2}+1)}\{4\beta(\alpha+\beta)^{4}[\beta^{2}(\alpha+\beta)-\alpha(1-\alpha\beta)]m_{Q}\nonumber\\&&{}
+(1-\alpha\beta)(\alpha^{2}+1)[(1-\alpha\beta)^{2}-\alpha\beta(\alpha+\beta)^{2}]m_{Q'}\},\nonumber
\end{eqnarray}
for $\Omega_{QQQ'}^{*}$, and
\begin{eqnarray}
\rho_{1}(s)&=&-\frac{3}{2^{4}\pi^{4}}\int_{\alpha_{min}}^{\alpha_{max}}d\alpha\{\int_{0}^{\beta_{1}}d\beta+\int_{\beta_{2}}^{\frac{1}{\alpha}}d\beta\}\frac{\alpha^{2}\beta^{2}(1-\alpha\beta)^{2}}{(\alpha+\beta)^{4}}[\alpha\beta(1-\alpha\beta)
s-((\alpha+\beta)^{2}m_{Q}^{2}+(1-\alpha\beta)m_{Q'}^{2})]s\nonumber\\&&{}
-\frac{3^{2}}{2^{5}\pi^{4}}\int_{\alpha_{min}}^{\alpha_{max}}d\alpha\{\int_{0}^{\beta_{1}}d\beta+\int_{\beta_{2}}^{\frac{1}{\alpha}}d\beta\}\frac{\alpha\beta(1-\alpha\beta)}{(\alpha+\beta)^{4}}[\alpha\beta(1-\alpha\beta)
s-((\alpha+\beta)^{2}m_{Q}^{2}+(1-\alpha\beta)m_{Q'}^{2})]^{2}\nonumber\\&&{}
-\frac{3}{2^{4}\pi^{4}}m_{Q}^{2}\int_{\alpha_{min}}^{\alpha_{max}}d\alpha\{\int_{0}^{\beta_{1}}d\beta+\int_{\beta_{2}}^{\frac{1}{\alpha}}d\beta\}\frac{\alpha\beta}{(\alpha+\beta)^{2}}[\alpha\beta(1-\alpha\beta)
s-((\alpha+\beta)^{2}m_{Q}^{2}+(1-\alpha\beta)m_{Q'}^{2})]\nonumber\\&&{}
+\frac{3\langle
g^{2}G^{2}\rangle}{2^{7}\pi^{4}}\int_{\alpha_{min}}^{\alpha_{max}}d\alpha\{\int_{0}^{\beta_{1}}d\beta+\int_{\beta_{2}}^{\frac{1}{\alpha}}d\beta\}\frac{\alpha^{2}\beta^{2}(1-\alpha\beta)}{(\alpha+\beta)^{2}},\nonumber\\
\rho_{2}(s)&=&-\frac{3}{2^{4}\pi^{4}}m_{Q'}\int_{\alpha_{min}}^{\alpha_{max}}d\alpha\{\int_{0}^{\beta_{1}}d\beta+\int_{\beta_{2}}^{\frac{1}{\alpha}}d\beta\}\frac{\alpha\beta(1-\alpha\beta)^{2}}{(\alpha+\beta)^{4}}[\alpha\beta(1-\alpha\beta)
s-((\alpha+\beta)^{2}m_{Q}^{2}+(1-\alpha\beta)m_{Q'}^{2})]s\nonumber\\&&{}
-\frac{3}{2^{4}\pi^{4}}m_{Q'}\int_{\alpha_{min}}^{\alpha_{max}}d\alpha\{\int_{0}^{\beta_{1}}d\beta+\int_{\beta_{2}}^{\frac{1}{\alpha}}d\beta\}\frac{1-\alpha\beta}{(\alpha+\beta)^{4}}[\alpha\beta(1-\alpha\beta)
s-((\alpha+\beta)^{2}m_{Q}^{2}+(1-\alpha\beta)m_{Q'}^{2})]^{2}\nonumber\\&&{}
-\frac{3}{2^{4}\pi^{4}}m_{Q}^{2}m_{Q'}\int_{\alpha_{min}}^{\alpha_{max}}d\alpha\{\int_{0}^{\beta_{1}}d\beta+\int_{\beta_{2}}^{\frac{1}{\alpha}}d\beta\}\frac{1}{(\alpha+\beta)^{2}}[\alpha\beta(1-\alpha\beta)
s-((\alpha+\beta)^{2}m_{Q}^{2}+(1-\alpha\beta)m_{Q'}^{2})]\nonumber\\&&{}
-\frac{\langle
g^{2}G^{2}\rangle}{2^{6}\pi^{4}}m_{Q'}\int_{\alpha_{min}}^{\alpha_{max}}d\alpha\{\int_{0}^{\beta_{1}}d\beta+\int_{\beta_{2}}^{\frac{1}{\alpha}}d\beta\}\frac{(1-\alpha\beta)[(1-\alpha\beta)^{2}+\alpha\beta(\alpha+\beta)^{2}]}{(\alpha+\beta)^{4}},\nonumber
\end{eqnarray}
for $\Omega'_{QQQ'}$. The integration limits are given by
\begin{eqnarray}
\alpha_{min}&=&\sqrt{\frac{(s^{2}-4m_{Q}^{2}s-2m_{Q'}^{2}s-4m_{Q}^{2}m_{Q'}^{2}+m_{Q'}^{4})-(s-m_{Q'}^{2})\sqrt{[s-(2m_{Q}-m_{Q'})^{2}][s-(2m_{Q}+m_{Q'})^{2}]}}{8m_{Q}^{2}s}},\nonumber\\
\alpha_{max}&=&\sqrt{\frac{(s^{2}-4m_{Q}^{2}s-2m_{Q'}^{2}s-4m_{Q}^{2}m_{Q'}^{2}+m_{Q'}^{4})+(s-m_{Q'}^{2})\sqrt{[s-(2m_{Q}-m_{Q'})^{2}][s-(2m_{Q}+m_{Q'})^{2}]}}{8m_{Q}^{2}s}},\nonumber\\
\beta_{1}&=&\frac{\alpha(s-2m_{Q}^{2}+m_{Q'}^{2})-\sqrt{\alpha^{2}[s-(2m_{Q}^{2}-m_{Q'}^{2})]^{2}-4(\alpha^{2}m_{Q}^{2}+m_{Q'}^{2})(m_{Q}^{2}+\alpha^{2}s)}}{2(\alpha^{2}
s+m_{Q}^{2})}, \mbox{and} \nonumber\\
\beta_{2}&=&\frac{\alpha(s-2m_{Q}^{2}+m_{Q'}^{2})+\sqrt{\alpha^{2}[s-(2m_{Q}^{2}-m_{Q'}^{2})]^{2}-4(\alpha^{2}m_{Q}^{2}+m_{Q'}^{2})(m_{Q}^{2}+\alpha^{2}s)}}{2(\alpha^{2}
s+m_{Q}^{2})}.\nonumber
\end{eqnarray}

\section{Numerical analysis}\label{sec3}
In numerical analysis, not both but only the sum rule (\ref{sum rule
q}) will be numerically analyzed for brevity, and the input values
are taken as $m_{c}=1.25~\mbox{GeV}, m_{b}=4.20~\mbox{GeV}$ with
$\langle g^{2}G^{2}\rangle=0.88~\mbox{GeV}^{4}$. Complying with the
standard procedure of sum rule method, the threshold $s_{0}$ and
Borel parameter $M^{2}$ are varied to find the optimal stability
window, in which the perturbative contribution should be larger than
the condensate contributions while the pole contribution larger than
continuum contribution. Accordingly, the sum rule windows are taken
as $\sqrt{s_0}=5.5\sim6.5~\mbox{GeV}$,
$M^{2}=6.5\sim8.0~\mbox{GeV}^{2}$ for $\Omega_{ccc}$,
$\sqrt{s_0}=14.0\sim15.0~\mbox{GeV}$,
$M^{2}=14.0\sim15.5~\mbox{GeV}^{2}$ for $\Omega_{bbb}$,
$\sqrt{s_0}=8.0\sim9.0~\mbox{GeV}$,
$M^{2}=8.5\sim10.0~\mbox{GeV}^{2}$ for $\Omega_{ccb}$,
$\sqrt{s_0}=8.5\sim9.5~\mbox{GeV}$,
$M^{2}=8.5\sim10.0~\mbox{GeV}^{2}$ for $\Omega_{ccb}^{*}$,
$\sqrt{s_0}=11.0\sim12.0~\mbox{GeV}$,
$M^{2}=10.0\sim11.5~\mbox{GeV}^{2}$ for $\Omega_{bbc}$,
$\sqrt{s_0}=11.5\sim12.5~\mbox{GeV}$,
$M^{2}=10.0\sim11.5~\mbox{GeV}^{2}$ for $\Omega_{bbc}^{*}$,
$\sqrt{s_0}=8.5\sim9.5~\mbox{GeV}$,
$M^{2}=8.5\sim10.0~\mbox{GeV}^{2}$ for $\Omega_{ccb}^{'}$, and
$\sqrt{s_0}=11.5\sim12.5~\mbox{GeV}$,
$M^{2}=10.0\sim11.5~\mbox{GeV}^{2}$ for $\Omega_{bbc}^{'}$,
respectively. The corresponding Borel curves are shown in Figs. 1-4.
In Table \ref{table:2}, the numerical results are presented,
together with the predictions from other theoretical approaches. It
is worth noting that uncertainty in our results are merely owing to
the sum rule windows, not involving the ones rooting in the
variation of the quark masses and QCD parameters. Note that the QCD
$O(\alpha_s)$ corrections are not included in this work, whose
calculations for triply heavy baryons are quite complicated and
tedious as one has to tackle some three-loop massive propagator
diagrams. However, it is expected that the QCD $O(\alpha_s)$
corrections might be under control since a partial cancellation
occurs in the ratio obtaining the mass sum rules (\ref{sum rule q})
and (\ref{sum rule m}). This has been proved to be true in the
analysis for the singly heavy baryons (the radiative corrections to
the perturbative terms increase the calculated baryon masses by
about $10\%$) \cite{alfa} and for the heavy mesons (the value of
$f_{D}$ increases by $12\%$ after the inclusion of the $O(\alpha_s)$
correction) \cite{Narison}. After a detailed comparison, one can
find that the central values of our results are lower than other
predictions from potential models, in particular, for
$\Omega_{bbb}$, slightly more than $\mbox{1~GeV}$, whereas the
relative discrepancy approximates to $10\%$. In addition, for
$\Omega_{ccc}$, our result is in good agreement with the lattice QCD
simulation in Ref. \cite{lat}, but the other comparisons can not be
made for lack of lattice results at present.

\begin{figure}
\centerline{\epsfysize=4.2truecm
\epsfbox{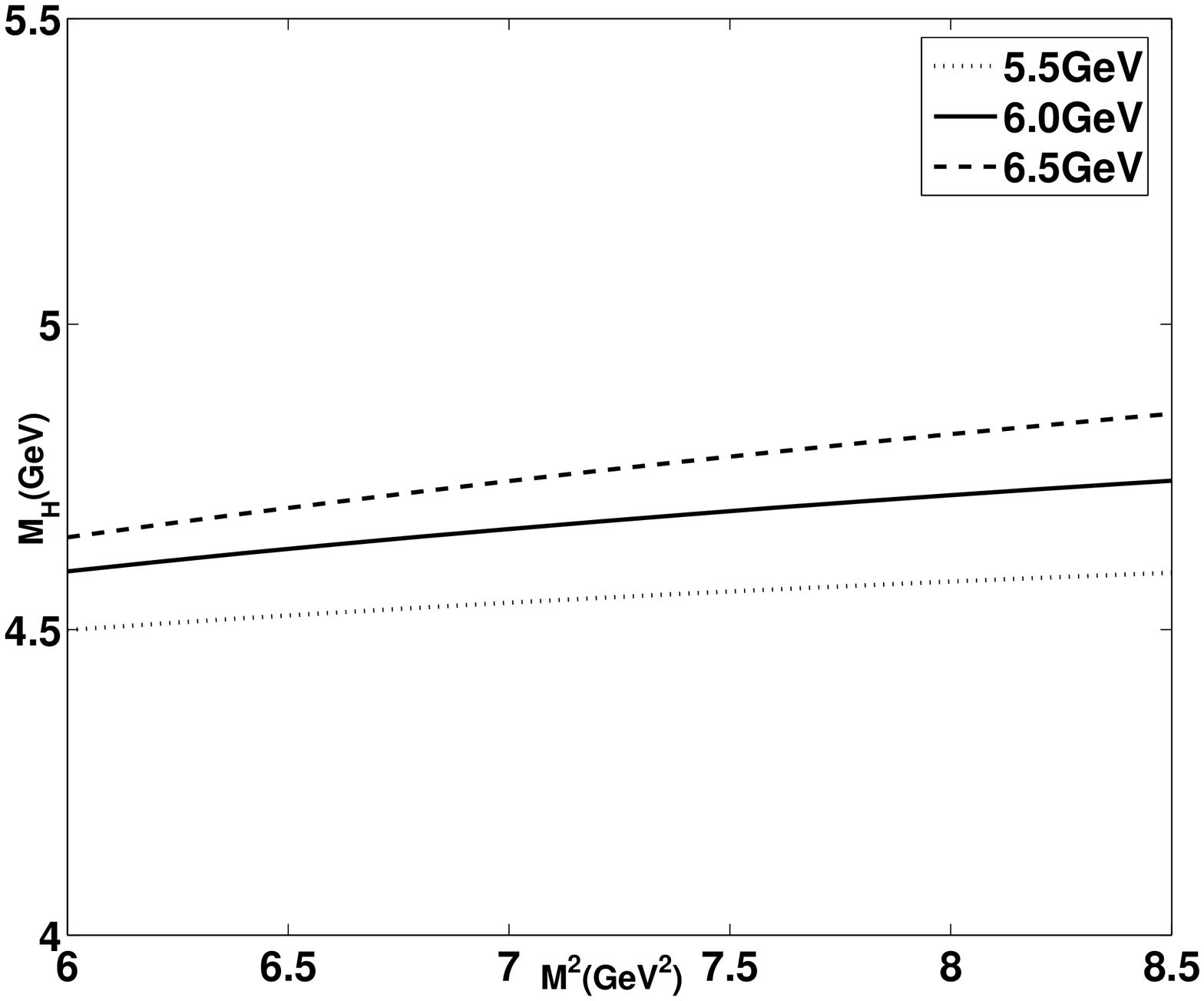}\epsfysize=4.2truecm \epsfbox{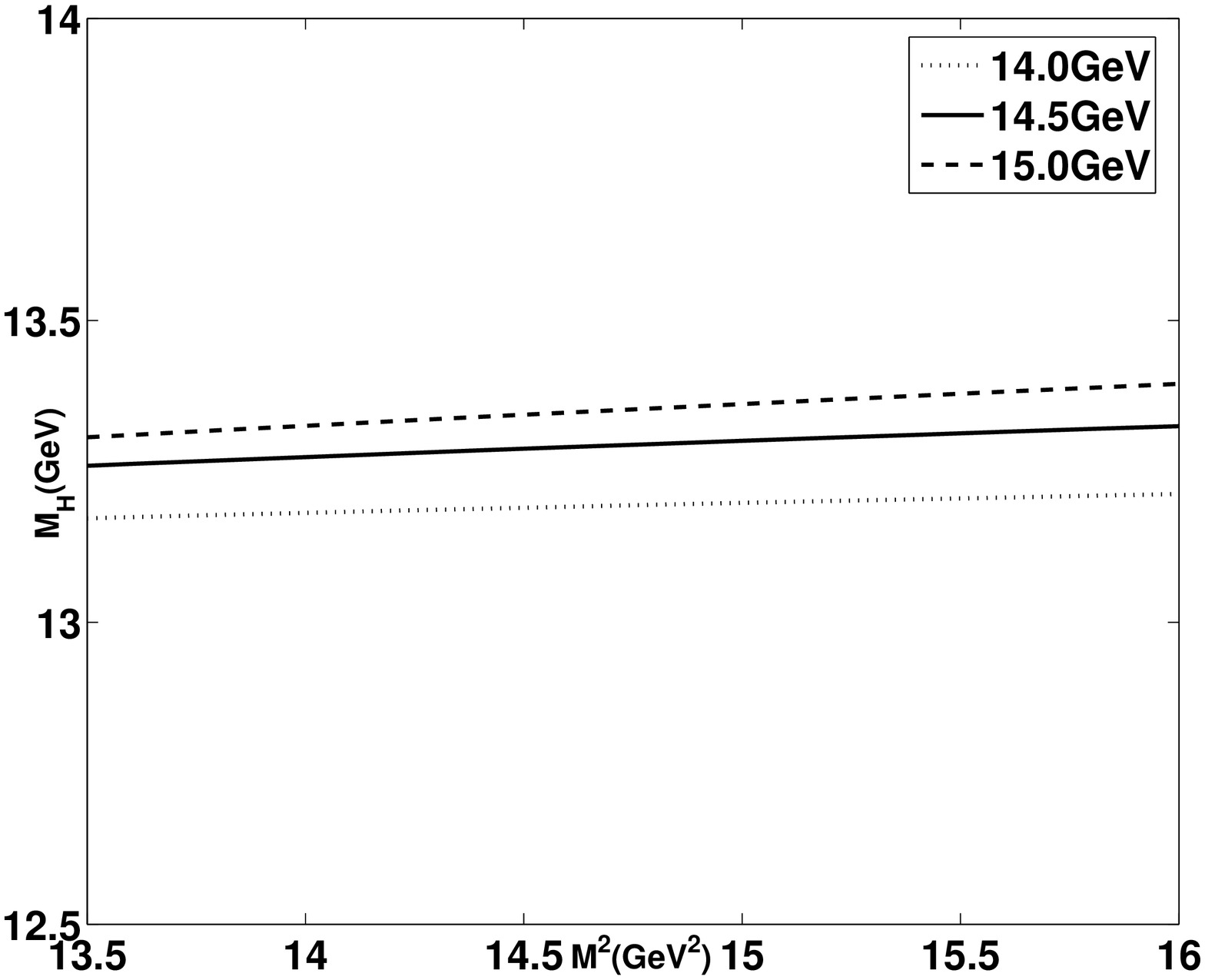}}\caption{The
dependence on $M^2$ for the masses of $\Omega_{ccc}$ and
$\Omega_{bbb}$ from sum rule (\ref{sum rule q}). The continuum
thresholds are taken as $\sqrt{s_0}=5.5\sim6.5~\mbox{GeV}$ and
$\sqrt{s_0}=14.0\sim15.0~\mbox{GeV}$.} \label{fig:1}
\end{figure}

\begin{figure}
\centerline{\epsfysize=4.2truecm
\epsfbox{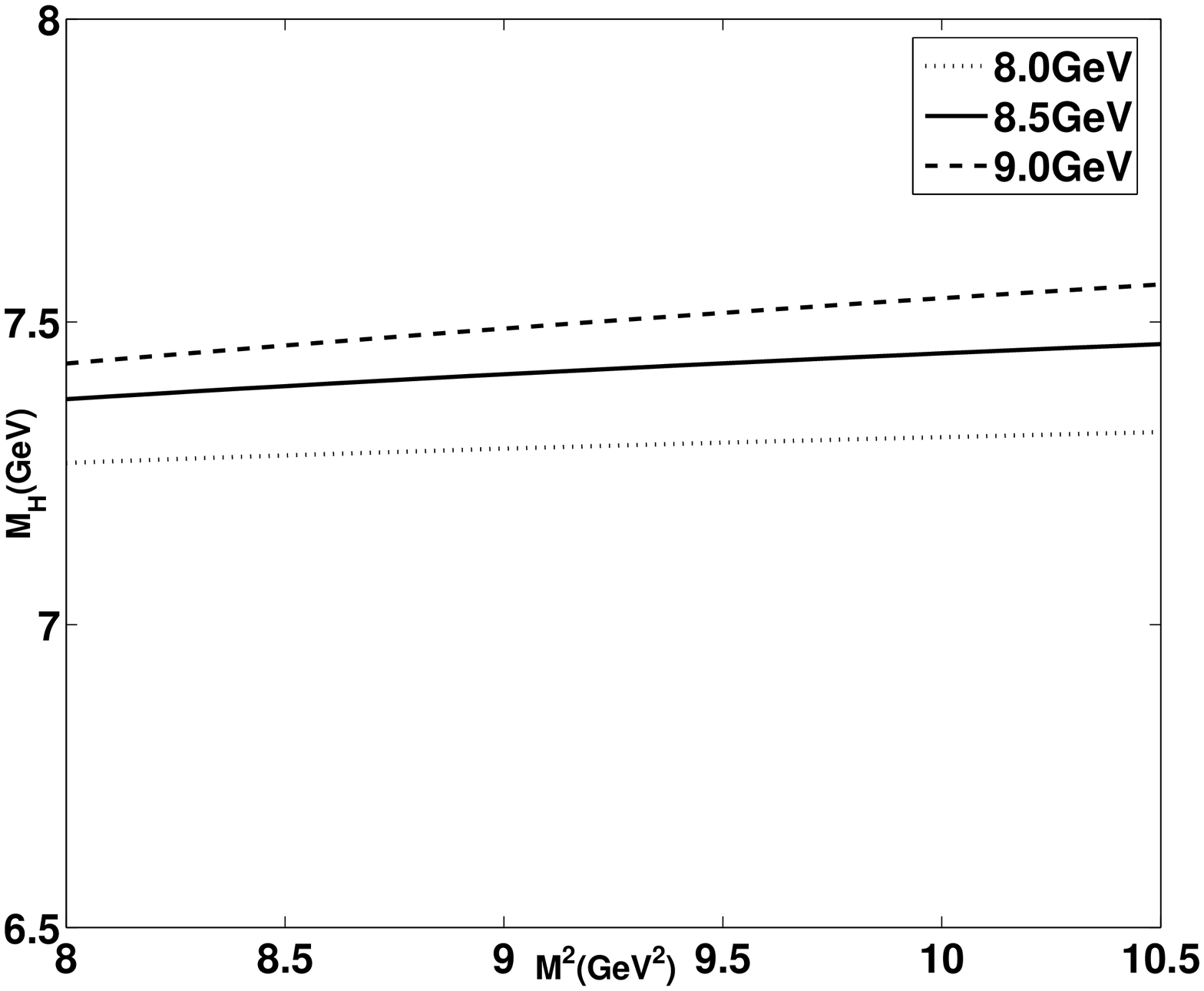}\epsfysize=4.2truecm
\epsfbox{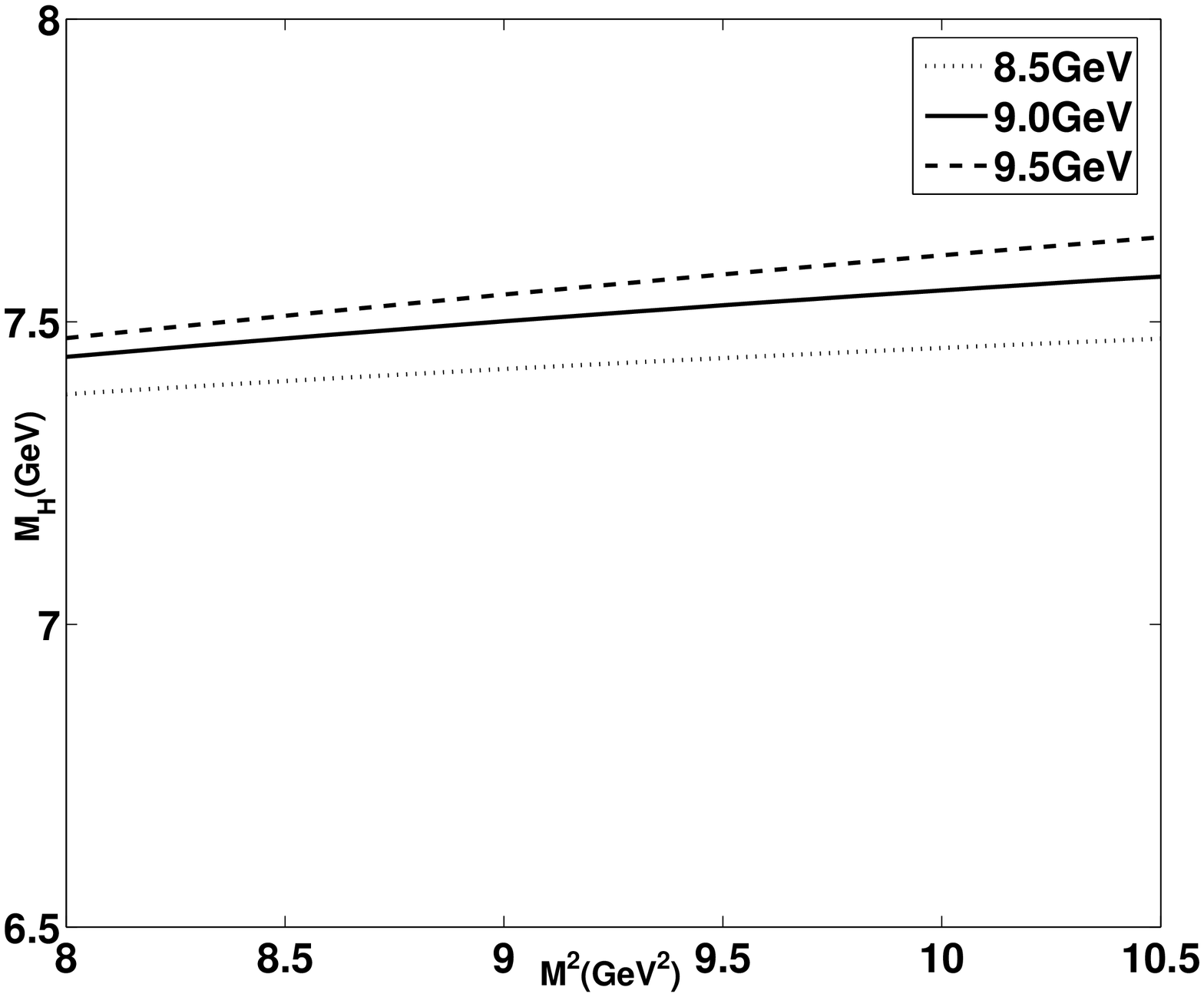}}\caption{The dependence on $M^2$ for the
masses of $\Omega_{ccb}$ and $\Omega_{ccb}^{*}$ from sum rule
(\ref{sum rule q}). The continuum thresholds are taken as
$\sqrt{s_0}=8.0\sim9.0~\mbox{GeV}$ and
$\sqrt{s_0}=8.5\sim9.5~\mbox{GeV}$.} \label{fig:2}
\end{figure}

\begin{figure}
\centerline{\epsfysize=4.2truecm
\epsfbox{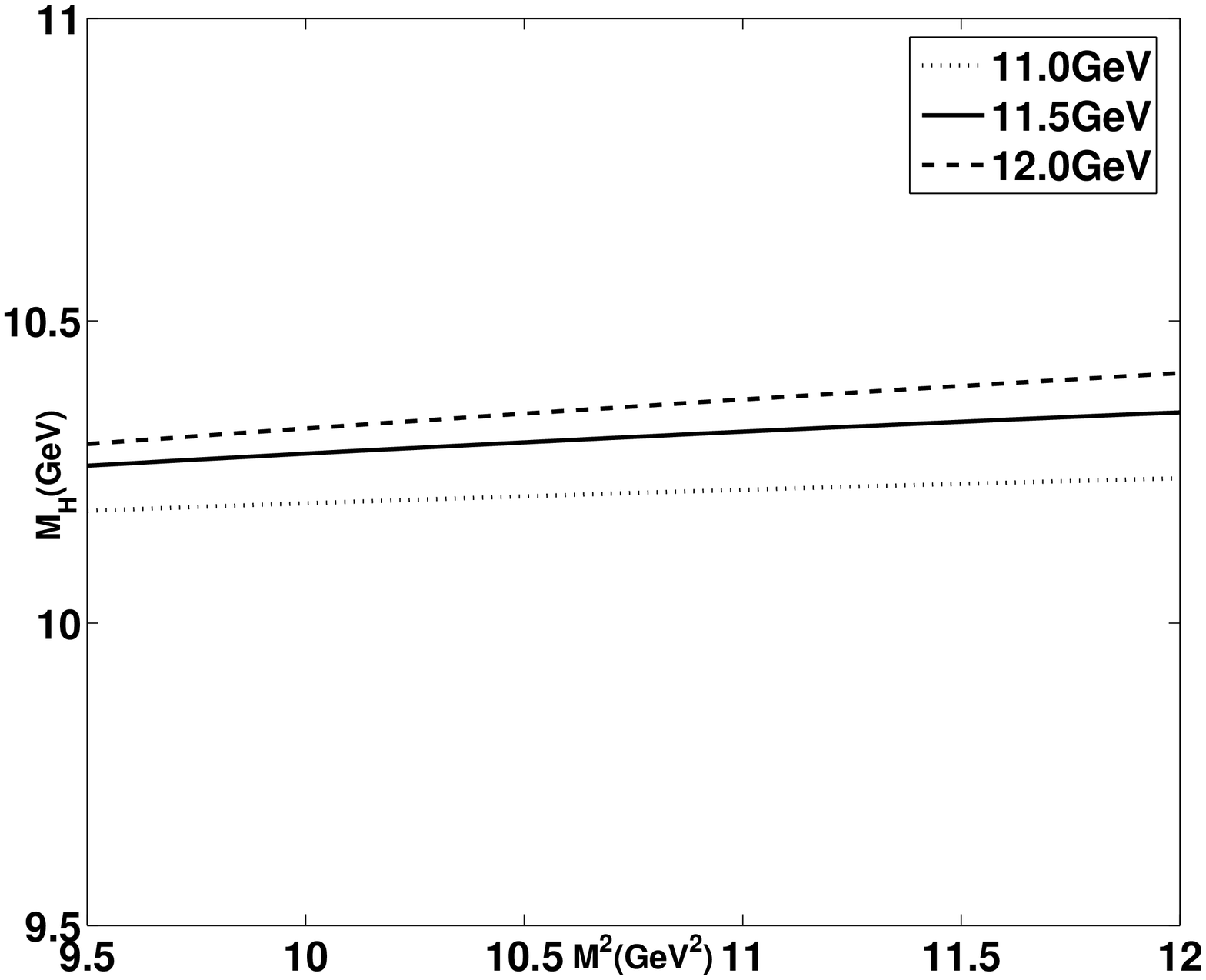}\epsfysize=4.2truecm
\epsfbox{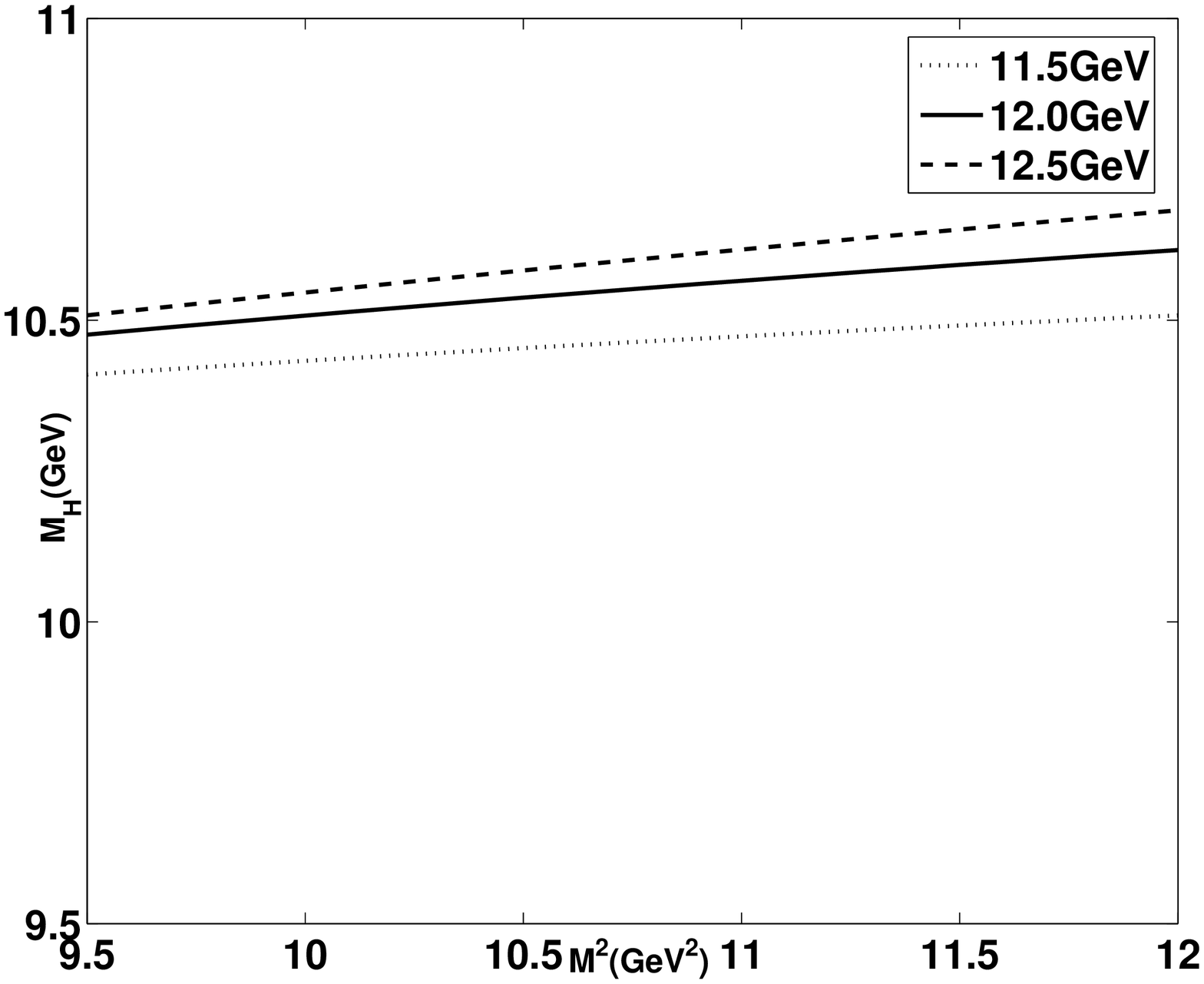}}\caption{The dependence on $M^2$ for the
masses of $\Omega_{bbc}$ and $\Omega_{bbc}^{*}$ from sum rule
(\ref{sum rule q}). The continuum thresholds are taken as
$\sqrt{s_0}=11.0\sim12.0~\mbox{GeV}$ and
$\sqrt{s_0}=11.5\sim12.5~\mbox{GeV}$.} \label{fig:3}
\end{figure}

\begin{figure}
\centerline{\epsfysize=4.2truecm
\epsfbox{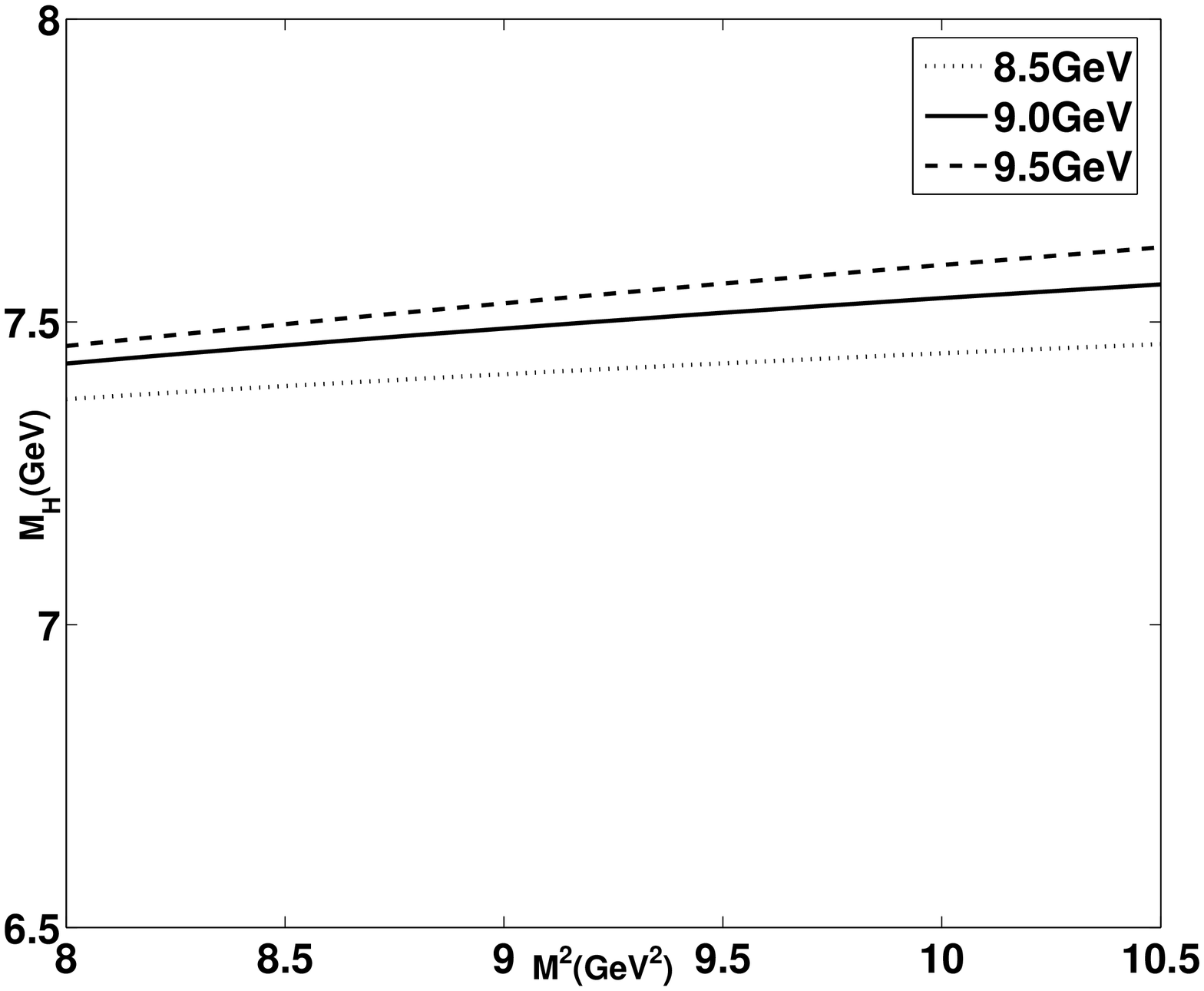}\epsfysize=4.2truecm
\epsfbox{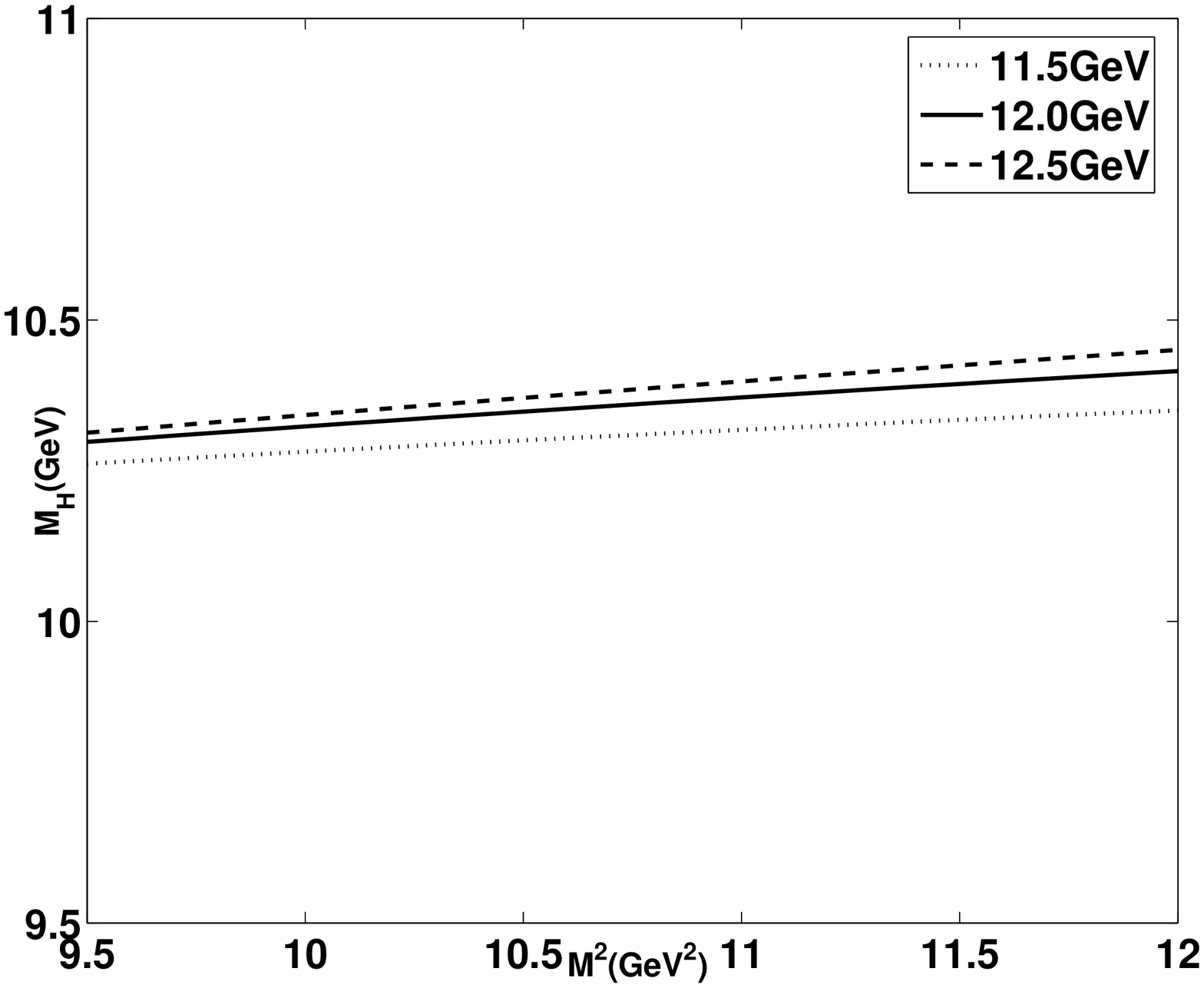}}\caption{The dependence on $M^2$ for the masses
of $\Omega_{ccb}^{'}$ and $\Omega_{bbc}^{'}$ from sum rule (\ref{sum
rule q}). The continuum thresholds are taken as
$\sqrt{s_0}=8.5\sim9.5~\mbox{GeV}$ and
$\sqrt{s_0}=11.5\sim12.5~\mbox{GeV}$.} \label{fig:4}
\end{figure}

\begin{table}[htb!]\caption{ The mass spectra of triply heavy baryons (mass in
unit of$~\mbox{GeV}$).}
 \centerline{\begin{tabular}{ c c c c c c c c c c c c c}  \hline\hline
Baryon                          & quark content          &     $J^{P}$            &  $S_{d}$     &  $L_{d}$     &   $J_{d}^{P_{d}}$         &     This work       &Ref. \cite{Hasenfratz} &Ref. \cite{Bjorken} &Ref. \cite{lat}      &Ref. \cite{YuJia}   &Ref. \cite{renewed1} &Ref. \cite{Martynenko}\\
\hline
 $\Omega_{ccc}$                 &$\{cc\}c$               &  $\frac{3}{2}^{+}$     &     1        &      0       &        $1^{+}$            &   $4.67\pm0.15$     &    $4.79$             &    $4.925$         &     $4.681$         &   $4.76$           &   $4.777$           &   $4.803$           \\
\hline
 $\Omega_{bbb}$                 &$\{bb\}b$               &  $\frac{3}{2}^{+}$     &     1        &      0       &        $1^{+}$            &   $13.28\pm0.10$    &    $14.30$            &    $14.760$        &                     &   $14.37$          &   $14.276$          &   $14.569$          \\
 \hline
 $\Omega_{ccb}$                 &$\{cc\}b$               &  $\frac{1}{2}^{+}$     &     1        &      0       &        $1^{+}$            &   $7.41\pm0.13$     &                       &                    &                     &                    &   $7.984$           &   $8.018$           \\
\hline
 $\Omega_{ccb}^{*}$             &$\{cc\}b$               &  $\frac{3}{2}^{+}$     &     1        &      0       &        $1^{+}$            &   $7.45\pm0.16$     &    $8.03$             &    $8.200$         &                     &   $7.98$           &   $8.005$           &   $8.025$           \\
\hline
 $\Omega_{bbc}$                 &$\{bb\}c$               &  $\frac{1}{2}^{+}$     &     1        &      0       &        $1^{+}$            &   $10.30\pm0.10$    &                       &                    &                     &                    &   $11.139$          &   $11.280$          \\
\hline
 $\Omega_{bbc}^{*}$             &$\{bb\}c$               &  $\frac{3}{2}^{+}$     &     1        &      0       &        $1^{+}$            &   $10.54\pm0.11$    &    $11.20$            &    $11.480$        &                     &   $11.19$          &   $11.163$          &   $11.287$          \\
 \hline
 $\Omega_{ccb}^{'}$             &$[cc]b$                 &  $\frac{1}{2}^{+}$     &     0        &      0       &        $0^{+}$            &   $7.49\pm0.10$     &                       &                    &                     &                    &                     &                     \\
\hline
 $\Omega_{bbc}^{'}$             &$[bb]c$                 &  $\frac{1}{2}^{+}$     &     0        &      0       &        $0^{+}$            &   $10.35\pm0.07$    &                       &                    &                     &                    &                     &                     \\
\hline\hline
\end{tabular}}
\label{table:2}
\end{table}

\section{Summary and outlook}\label{sec4}
In a tentative $(QQ)-(Q')$ configuration, the QCD sum rules have
been employed to calculate the masses of triply heavy baryon $QQQ'$
($Q=Q'$ or $Q\neq Q'$), including the contributions of the operators
up to dimension six in OPE. The mass values extracted from the sum
rules are collected in comparison with other theoretical
predictions. The results in this work are lower than the predictions
from potential models, nevertheless, the one for $\Omega_{ccc}$ is
well compatible with the existing lattice study. Indubitably, there
are still plenty of problems desiderated to resolve. Experimentally,
the evidence on triply heavy baryons are expected to reveal nature
of them, especially after the putting into operation of the Large
Hadron Collider. In theory, in order to improve on the accuracy of
the QCD sum rule analysis for triply heavy baryons, one certainly
needs to take into account the QCD $O(\alpha_s)$ corrections to the
sum rules in the further work. Additionally, it may be needed to
carry out a comprehensive study on triply heavy baryon spectra from
lattice QCD stimulations for the future.

\begin{acknowledgments}
This work was supported in part by the National Natural Science
Foundation of China under Contract No.10675167.
\end{acknowledgments}


\begin{thebibliography}{99}

\bibitem{Hasenfratz}P.~Hasenfratz, R.~R.~Horgan, J.~Kuti, and J.~M.~Richard, Phys. Lett. B {\bf 94}, 401 (1980).

\bibitem{Bjorken}J.~D.~Bjorken, Preprint FERMILAB-Conf-85-069.

\bibitem{Bagan}E.~Bagan, M.~Chabab, H.~G.~Dosch, and S.~Narison, Phys. Lett. B {\bf 278}, 367
(1992); B {\bf 287}, 176 (1992); B {\bf 301}, 243 (1993); E.~Bagan,
M.~Chabab, and S.~Narison, Phys. Lett. B {\bf 306}, 350 (1993).

\bibitem{heavy baryons}Y.~B.~Dai, C.~S.~Huang, C.~Liu, and C.~D.~L\"{u}, Phys. Lett. B {\bf371}, 99 (1996);
F.~O.~Dur\~{a}es and M.~Nielsen, Phys. Lett. B {\bf 658}, 40 (2007);
Z.~G.~Wang, Eur. Phys. J. C {\bf 54}, 231 (2008); J.~R.~Zhang and
M.~Q.~Huang, Phys. Rev. D {\bf 77}, 094002 (2008); D {\bf 78},
094007 (2008); D {\bf 78}, 094015 (2008).

\bibitem{EFT}N.~Brambilla, A.~Vairo, and T.~R\"{o}sch, Phys. Rev. D {\bf 72}, 034021 (2005).

\bibitem{lat}T.~W.~Chiu and T.~H.~Hsieh, Nucl. Phys. A {\bf 755},
471c (2005).

\bibitem{YuJia}Y.~Jia, JHEP {\bf 0610}, 073 (2006).


\bibitem{renewed1}A.~Berotas and V.~\v{S}imonis, arXiv:0808.1220.

\bibitem{quark model}J.~Vijande, H.~Garcilazo, A.~Valcarce, and F.~Fern\'{a}ndez, Phys.
Rev. D {\bf 70}, 054022 (2004).

\bibitem{Martynenko}A.~P.~Martynenko, Phys. Lett. B {\bf 663}, 317 (2008).

\bibitem{renewed2}B.~Patel, A.~Majethiya, and P.~C.~Vinodkumar, arXiv:0808.2880.


\bibitem{overview1}E.~S.~Swanson, Phys. Rep. {\bf 429}, 243 (2006).

\bibitem{overview2}Particle Data Group, C.~Amsler et al., Phys. Lett. B {\bf 667}, 1 (2008).

\bibitem{production1}M.~A.~Gomshi Nobary, Phys. Lett. B {\bf 559},
239 (2003).

\bibitem{production2}M.~A.~Gomshi Nobary and R.~Sepahvand, Phys. Rev. D {\bf 71}, 034024
(2005); Nucl. Phys. {\bf B741}, 34 (2006); Phys. Rev. D {\bf 76},
114006 (2007).

\bibitem{production3}M.~A.~Gomshi Nobary, B.~Nikoobakht, and J.~Naji, Nucl. Phys. A {\bf 789}, 243 (2007).

\bibitem{svzsum}M.~A.~Shifman, A.~I.~Vainshtein, and V.~I.~Zakharov, Nucl. Phys. {\bf B147}, 385 (1979); {\bf B147}, 448 (1979);
 V.~A.~Novikov, M.~A.~Shifman, A.~I.~Vainshtein, and V.~I.~Zakharov, Fortschr. Phys. {\bf 32}, 585 (1984).

\bibitem{charmonium and bottomonium}L.~J.~Reinders, H.~R.~Rubinstein, and S.~Yazaki, Nucl.
Phys. {\bf B186}, 109 (1981); Phys. Rep. {\bf 127}, 1 (1985).


\bibitem{c and b}C.~A.~Dominguez, G.~R.~Gluckman, and N.~Paver, Phys. Lett. B {\bf 333}, 184 (1994);
S.~Narison, Phys. Lett. B {\bf 341}, 73 (1994).

\bibitem{alpha2}M.~Eidem\"{u}ller and M.~Jamin, Phys. Lett. B {\bf 498}, 203 (2001);
G.~Corcella and A.~H.~Hoang, Phys. Lett. B {\bf 554}, 133 (2003);
K.~G.~Chetyrkin, J.~H.~Kuhn, and M.~Steinhauser, Nucl. Phys. {\bf
B505}, 40 (1997).

\bibitem{Bc}E.~Bagan et al., Z. Phys. C {\bf 64}, 57 (1994); P.~Colangelo, G.~Nardulli, and N.~Paver, Z. Phys. C {\bf 57}, 43 (1993);
V.~V.~Kiselev, A.~K.~Likhoded, and A.~I.~Onishchenko, Nucl. Phys.
{\bf B569}, 473 (2000).

\bibitem{evs}E.~V.~Shuryak, Nucl. Phys. {\bf B198}, 83 (1982).

\bibitem{Ioffe}B.~L.~Ioffe, Nucl. Phys. {\bf B188}, 317 (1981); {\bf B191}, 591(E) (1981); Z. Phys. C {\bf 18}, 67 (1983).

\bibitem{alfa}S.~Groote, J.~G.~K\"{o}rner, and O.~I.~Yakovlev, Phys. Rev. D
{\bf 55}, 3016 (1997).

\bibitem{Narison}S.~Narison, Phys. Lett. B {\bf 605}, 319 (2005).

\end{thebibliography}
\end{document}